\documentstyle[aps,prb,epsf, multicol]{revtex}
\tighten
\begin{document}
\draft
\newcommand{\bn}{{\bf n}}
\newcommand{\bp}{{\bf p}}
\newcommand{\br}{{\bf r}}
\newcommand{\bq}{{\bf q}}
\newcommand{\bj}{{\bf j}}
\newcommand{\bE}{{\bf E}}
\newcommand{\eps}{\varepsilon}
\newcommand{\la}{\langle}
\newcommand{\ra}{\rangle}
\newcommand{\cK}{{\cal K}}
\newcommand{\cD}{{\cal D}}
\newcommand{\mybeginwide}{
    \end{multicols}\widetext
    \vspace*{-0.2truein}\noindent
    \hrulefill\hspace*{3.6truein}
}
\newcommand{\myendwide}{
    \hspace*{3.6truein}\noindent\hrulefill
    \begin{multicols}{2}\narrowtext\noindent
}

\title{
  Cascade Boltzmann - Langevin approach to higher-order current
  correlations in diffusive metal contacts
}
\author{K.\ E.\ Nagaev
}
\address{
  Institute of Radioengineering and Electronics,
  Russian Academy of Sciences, Mokhovaya ulica 11, 101999 Moscow,
  Russia\\}
\date\today
\maketitle
\bigskip
\begin{abstract}
The Boltzmann - Langevin approach is extended to calculations of
third and fourth cumulants of current in diffusive-metal
contacts. These cumulants result from indirect correlations
between current fluctuations, which may be considered as "noise of
noise". The calculated third cumulant coincides exactly with its
quantum-mechanical value. The fourth cumulant tends to its
quantum-mechanical value $-e^3I/105$ at high voltages and to a
positive  value $2e^2T/3R$ at $V=0$ changing its sign at $eV \sim
20T$.
\end{abstract}
\begin{multicols}{2}
\narrowtext

The fluctuation - dissipation theorem relates the spectral density
of noise in equilibrium with the corresponding linear response of
the system. This is why the second cumulant of current presents
new information about the system only under nonequilibrium
conditions. Nonequilibrium noise in mesoscopic systems has been
extensively studied in the past decade.\cite{Blanter-00a} Unlike
the second cumulant of current, its higher cumulants are not
related with any average characteristics of the system even in
equilibrium. Hence these quantities may be also of experimental
interest.

 In recent years, higher-order
correlations of currents in mesoscopic systems received
considerable attention of theorists. This work was pioneered by
Levitov and Lesovik,\cite{Levitov-93} who calculated the
distribution of charge transmitted through a single-channel
quantum contact. Subsequently, these calculations were extended to
phase-coherent diffusive contacts in the zero-temperature
limit.\cite{Lee-95} Very recently, the third cumulant of current
was calculated for quasi-one-dimensional diffusive-metal contacts
for arbitrary temperatures and voltages\cite{Gutman-02} using the
nonlinear $\sigma$-model and Keldysh formalism.\cite{Kamenev-99}

Along with the fully quantum-mechanical approach, higher-order
correlations have been studied using a semiclassical description,
which ignores quantum interference of wave functions yet takes
into account the Fermi statistics of electrons. Semiclassical
calculations of higher-order cumulants were performed for
double-barrier structures\cite{deJong-96} using a master equation
and for chaotic cavities\cite{Blanter-00b} based on the minimal
correlation approach.\cite{Blanter-99}

A semiclassical Boltzmann - Langevin approach\cite{Kogan-69} has
been very successful in describing the second cumulant of current
in diffusive contacts.\cite{Nagaev-92} In particular, it gives the
same value of the shot noise $S_I = 2eI/3$ as a fully
quantum-mechanical treatment.\cite{Beenakker-92} This is not
surprising because the corrections to semiclassical values from
quantum-interference effects like weak localization and mesoscopic
fluctuations of conductance in dirty metals contain a small
parameter\cite{Abrikosov} $1/p_F l_{imp}$, and the absence of this
parameter in the noise magnitude implies that quantum interference
is irrelevant to shot noise. As the values of higher cumulants
obtained by Lee et al.\cite{Lee-95} either do not contain the
quantum-interference prefactor as compared to the second cumulant,
they can be also obtained semiclassically. This conclusion is
supported by recent numerical results,\cite{Roche-02} which were
obtained in a semiclassical model with exclusion principle.

The present paper extends the semiclassical Boltzmann - Langevin
approach to calculations of the third and fourth cumulants of current
for diffusive-metal contacts.  We show that the main contribution to
these quantities still results from second-order correlation functions
of the Langevin sources in the Boltzmann - Langevin equation, whereas
the effect of higher-order correlations of these sources is negligibly
small in the diffusive case. As the result involves
several hierarchically coupled second-order correlators, this approach
may be termed "cascade".

The Boltzmann-Langevin equation for fluctuations reads
$$
  \left[
     \frac{\partial}{\partial t}
     +
     {\bf v}\frac{\partial}{\partial{\bf r}}
     +
     e{\bf Ev}\frac{\partial}{\partial \eps}
  \right] \,
  \delta f(\bp,\br,t)
  +
  \delta I
$$ \begin{equation}
  =
  -e\,\delta{\bf E\,v} \frac{\partial f}{\partial\eps}
  +
  \delta J^{ext},
  \label{G-BL}
  \end{equation}
where $\delta I$ is the linearized collision integral and $\delta
J^{ext}$ is the random extraneous flux. Fluctuations of physical
quantities are expressed in terms of the fluctuations of the
distribution function $\delta f$. For example, the fluctuation of
current density is given by an expression
\begin{equation}
 \delta\bj(\br, t)
 =
 e\sum\limits_{\bp}
 {\bf v}
 \delta f(\bp, \br, t).
 \label{G-dj}
\end{equation}
In principle, the $n$th-order correlation function of any physical
quantity may be calculated provided that the correlation functions
of the extraneous sources $\delta J^{ext}$ are known up to $n$th
order.

Kogan and Shulman\cite{Kogan-69} calculated the second-order
correlation function of the extraneous sources based on very
simple physical considerations. The collision integral in the
Boltzmann equation is of the form
\begin{equation}
 I(\bp, \br, t)
 =
 \sum\limits_{\bp'}
 \left[
   J(\bp\to\bp')
   -
   J(\bp'\to\bp)
 \right],
 \label{G-I_coll}
\end{equation}
where $J(\bp\to\bp')$ and $J(\bp'\to\bp)$ are the outgoing and
incoming scattering fluxes from and to state $\bp$. It is natural
to assume that all scattering events are local in space and the
scattering fluxes between different pairs of states are
statistically independent. Furthermore, since scattering events at
different instants of time are independent, the scattering between
each pair of states presents a Poissonian process whose
second-order correlation function $\la\delta J(t_1)\delta
J(t_2)\ra$ is proportional to its average rate
\begin{equation}
 J(\bp\to\bp')
 =
 W(\bp, \bp')
 f(\bp,\br, t)[1 - f(\bp', \br, t)],
\label{G-flux}
\end{equation}
where $W(\bp, \bp')$ is the classical scattering probability and
the factor $1 - f$ takes into account the Pauli principle. Hence
the correlation function of extraneous sources related with
randomness of scattering may be written in a form\cite{Kogan-69}
$$
 \la
   \delta J^{ext}(\bp_1,\br_1,t_1)
   \delta J^{ext}(\bp_2,\br_2,t_2)
 \ra
 =
 \delta(\br_1 - \br_2)
 \delta(t_1 - t_2)
$$ $$
 \times
 \Biggl\{
   \delta_{\bp_1\bp_2}
   \sum\limits_{\bp'}
   [
    J(\bp_1 \rightarrow \bp' )
    +
    J(\bp'  \rightarrow \bp_1)
   ]
   -
   J(\bp_1 \rightarrow \bp_2)
$$
\begin{equation}
   -
   J(\bp_2 \rightarrow \bp_1)
 \Biggr\}.
 \label{G-dJ^ext-2}
\end{equation}
The correlation function is positive if $\bp_1 = \bp_2$ and
negative if $\bp_1 \ne \bp_2$. In the latter case the only
correlation is possible through scattering from $\bp_1$ to $\bp_2$
and vice versa, but this results in fluctuations of the
corresponding occupancies of opposite signs.

In the case of a diffusive metal and low frequencies, the
Boltzmann - Langevin scheme is greatly simplified. It is
convenient to introduce a fluctuation of electric potential
$\delta\phi$ and random extraneous
current\cite{Nagaev-98,Sukhorukov-98}
\begin{equation}
 \delta\bj^{ext}(\br, t)
 =
 e\tau
 \sum\limits_{\bp}
 {\bf v}\delta J^{ext}(\bp),
 \label{D-dj^ext}
\end{equation}
where $\tau$ is the elastic scattering time. The fluctuation of
current density is given by an expression
\begin{equation}
 \delta\bj
 =
 -\sigma\nabla\delta\phi
 +
 \delta\bj^{ext},
 \label{D-dj.vs.dphi}
\end{equation}
where $\sigma$ is the metal conductivity. Because pile-up of
charge is forbidden in the quasi-static limit, one easily obtains
from the condition of current conservation that
\begin{equation}
 \sigma\nabla^2\delta\phi
 =
 \nabla\bj^{ext}.
 \label{D-dphi}
\end{equation}
The correlation function of extraneous currents
$$
 \la
  \delta j_{\alpha}^{ext}(\br_1, t_1)
  \delta j_{\beta}^{ext}(\br_2, t_2)
 \ra
 =
 2\sigma\delta_{\alpha\beta}
 \delta(\br_1 - \br_2)
 \delta(t_1 - t_2)
$$ \begin{equation}
 \times
 \int d\eps\,
 f(\eps,\br_1)
 [1 - f(\eps,\br_1)]
 \label{D-dj^ext-2}
\end{equation}
is easily obtained from Eq. (\ref{G-dJ^ext-2}). Expressions
(\ref{D-dj.vs.dphi}) - (\ref{D-dj^ext-2}) form a complete set of
equations for calculating the second cumulant of current.

\begin{figure}[t]
\epsfxsize8cm \centerline{\epsffile{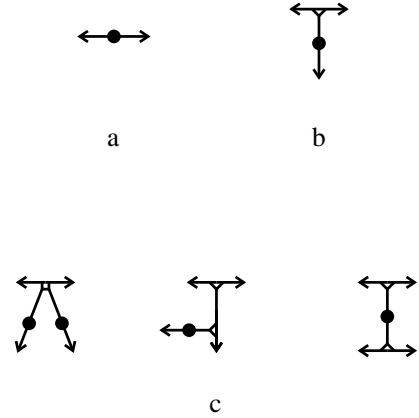}}
\caption{%
 Typical contributions to (a) second, (b) third, and (c) fourth
 cumulants of current in a diffusive metal. The arrows correspond
 to current fluctuations at different instances of time. Black
 circles correspond to the correlators of Langevin sources,
 empty triangles and squares, to their first and second functional
 derivatives.
} \label{fig_1}
\end{figure}
It is natural to use the same notion of independent scattering
events for determining the correlation functions of Langevin
sources of higher order. According to the ideas of Kogan and
Shulman, the Langevin source is a fluctuation of the difference
between the outgoing and incoming scattering fluxes into a given
state $(\bp, \br)$. Since each scattering flux between a pair of
states is assumed to be a Poissonian process, all its cumulants
are proportional to the average scattering rate $1/\tau$. Hence
the third- and fourth-order correlation functions of extraneous
currents (\ref{D-dj^ext}) are proportional to $\tau^3$ (see
Appendix for details). As will be shown below, the contribution
from these correlations is negligibly small as compared to another
mechanism related with second-order correlations of Langevin
sources. These correlations may be called indirect, or "cascade".
The point is that the second-order correlation function of
extraneous sources is a functional of the distribution function
$f$ and hence may also fluctuate as the current itself. One may
roughly imagine these fluctuations as fluctuations of Nyquist
noise of a resistor caused by fluctuations of its temperature. A
fluctuation $\delta f$ results in a fluctuation
$$
 \delta
 \la
  \delta j_{\alpha}^{ext}(\br_1, t_1)
  \delta j_{\beta}^{ext}(\br_2, t_2)
 \ra
 =
 2\sigma\delta_{\alpha\beta}
 \delta(\br_1 - \br_2)
 \delta(t_1 - t_2)
$$ $$
 \times
 \int d\eps\,
 [1 - 2f(\eps,\br_1)]
 \delta f(\eps,\br_1,t_1).
$$
Note that the time scale for the fluctuation $\delta f$ is set by a
characteristic relaxation time and hence they are slow as compared
to the duration of a scattering event. As $\delta f$ is also
correlated with fluctuations of other quantities, this "noise of
noise" results in higher-order correlations of currents. Hence even
Gaussian extraneous sources may produce non-Gaussian fluctuations.

In a symbolic form, the expression for indirect third-order
correlations may be written in a form
\begin{equation}
 \la
   \delta I_1 \delta I_2 \delta I_3
 \ra
 =
 P_{123}
 \left\{
   \frac{
     \delta
     \la
       \delta I_1 \delta I_2
     \ra
   }{
     \delta f_4
   }
   \la
     \delta f_4 \delta I_3
   \ra
 \right\},
 \label{D-dI-3}
\end{equation}
where $P_{123}$ denotes a summation over all inequivalent
permutations of indices $(123)$ and $\delta\la\ldots\ra/\delta
f_4$ denotes a functional derivative with respect to
$f(\eps_4,\br_4, t_4)$. The products imply a convolution over the
arguments of the distribution functions with repeating indices.
The {\it irreducible} fourth-order correlation function is
presented by three groups of terms
$$
 \la:
   \delta I_1 \delta I_2 \delta I_3 \delta I_4
 :\ra
 =
 P_{1234}
 \left\{
   \frac{
     \delta^2
     \la
       \delta I_1 \delta I_2
     \ra
   }{
     \delta f_5 \delta f_6
   }
   \la
     \delta f_5 \delta I_3
   \ra
   \la
     \delta f_6 \delta I_4
   \ra
 \right.
$$ $$
 \left.
   +
   \frac{
     \delta
     \la
       \delta I_1 \delta I_2
     \ra
   }{
     \delta f_5
   }
   \frac{
     \delta
     \la
       \delta f_5 \delta I_3
     \ra
   }{
     \delta f_6
   }
   \la
     \delta f_6 \delta I_4
   \ra
 \right.
$$ \begin{equation}
 \left.
   +
   \frac{
     \delta
     \la
       \delta I_1 \delta I_2
     \ra
   }{
     \delta f_5
   }
   \la
     \delta f_5 \delta f_6
   \ra
   \frac{
     \delta
     \la
       \delta I_3 \delta I_4
     \ra
   }{
     \delta f_6
   }
 \right\}.
 \label{D-dI-4}
\end{equation}
The corresponding contributions are schematically illustrated by
diagrams in Fig. 1.

To evaluate expressions (\ref{D-dI-3}) and (\ref{D-dI-4}), one
must know the second-order correlation function of fluctuations
$\delta f$. In the quasi-static limit, the Boltzmann - Langevin
equation may be easily transformed into a stochastic diffusion
equation for the fluctuation $\delta f(\eps, \br, t)$
\begin{equation}
 D\nabla^2\delta f(\eps, \br)
 =
 \nabla\delta{\bf F}^{ext},
 \label{D-df}
\end{equation}
where
\begin{equation}
  \delta{\bf F}^{ext}(\eps)
  =
  \frac{\tau}{N_F}
  \sum\limits_{\bp}
  {\bf v}\delta J^{ext}(\bp)
  \delta
  (
    \eps_{\bp} - \eps
  ),
  \label{D-dF^ext-def}
\end{equation}
so one obtains from (\ref{G-dJ^ext-2}) that
$$
 \la
   \delta F_{\alpha}^{ext}(\eps, \br)
   \delta F_{\beta }^{ext}(\eps', \br')
 \ra
 =
 2\frac{D}{N_F}
 \delta(\br - \br')
 \delta(t - t')
$$ \begin{equation}
 \delta(\eps - \eps')
 \delta_{\alpha\beta}
 f(\eps)[1 - f(\eps)],
 \label{D-dF^ext-2}
\end{equation}
$$
 \la
   \delta F_{\alpha}^{ext}(\eps, \br)
   \delta j_{\beta }^{ext}( \br')
 \ra
 =
 2eD
 \delta(\br - \br')
 \delta(t - t')
$$ \begin{equation}
 \times
 \delta_{\alpha\beta}
 f(\eps)[1 - f(\eps)],
 \label{D-<dF^ext_dj^ext>}
\end{equation}
where $D$ is the diffusion coefficient and $N_F$ is the Fermi
density of states.

Consider the case of a quasi-one-dimensional diffusive wire of
length $L$ connecting two massive electrodes with a voltage drop
$V$ across it. In this case, all quantities may be considered as
depending only on the longitudinal coordinate $x$ along the wire.
The solution of Eq. (\ref{D-df}) is of the form
$$
  \delta f(x)
  =
  \frac{1}{D S_0}
  \int\limits_0^L dx'
  K(x, x')
  \int d^2 r_{\perp}
  \delta F_x^{ext}(\eps, x, \br_{\perp} ),
$$ \begin{equation}
  K(x, x')
  =
  \theta(x - x')
  -
  x/L,
  \label{Q-df}
\end{equation}
where $S_0$ is the area of the cross-section of the wire and
$\br_{\perp}$ are the transverse coordinates. Because the pile-up
of charge in the contact is forbidden in the quasi-static limit,
the fluctuation of current is independent of $x$ and is obtained
by averaging the extraneous current over the contact
volume\cite{Nagaev-92}
\begin{equation}
 \delta I
 =
 \int\limits_0^L
 \frac{dx}{L}
 \int d^2 r_{\perp}
 \delta j_x^{ext}.
 \label{Q-dI}
\end{equation}
The low-frequency Fourier transform of the correlation function of
$\delta I(t_1)$ and $\delta I(t_2)$ is given by
$$
  \la
    \delta I(\omega_1) \delta I(\omega_2)
  \ra
  =
  \delta
  (
    \omega_1 + \omega_2
  )
$$ \begin{equation}
  \times
  \frac{4\pi}{RL}
  \int\limits_0^L dx
  \int d\eps
  f(\eps, x)[1 - f(\eps,x)],
  \label{Q-<dI1dI2>}
\end{equation}
where $R$ is the resistance of the contact. By giving the
distribution function a variance $\delta f(\eps,x)$, one easily
obtains the functional derivative of Eq. (\ref{Q-<dI1dI2>}) with
respect to $f(\eps, x, \omega_3)$ in the form
\begin{equation}
  \frac{
    \delta
    \la
      \delta I(\omega_1) \delta I(\omega_2)
    \ra
  }{
    \delta f(\eps, x, \omega_3)
  }
  =
  \delta
  (
    \omega_1 + \omega_2 - \omega_3
  )
  \frac{4\pi}{RL}
  [ 1 - 2f(\eps, x) ].
  \label{Q-d<dI1dI2>/df}
\end{equation}
The second functional derivative of the same quantity is just
$$
  \frac{
    \delta^2
    \la
      \delta I(\omega_1) \delta I(\omega_2)
    \ra
  }{
    \delta f(\eps_3, x_3, \omega_3)
    \delta f(\eps_4, x_4, \omega_4)
  }
  =
  -
  \delta
  (
    \omega_1 + \omega_2 - \omega_3 - \omega_4
  )
$$ \begin{equation}
  \times
  \delta( \eps_3 - \eps_4 )
  \delta( x_3 - x_4 )
  \frac{8\pi}{RL}.
  \label{Q-d^2/df^2}
\end{equation}
The right-hand side of Eq. (\ref{D-dI-3}) allows three
inequivalent permutations of indices. Hence the low-frequency
Fourier transform of the third cumulant of current
$$
 S_3(\omega_1, \omega_2)
 =
 \int d(t_1 - t_3)
 \int d(t_2 - t_3)
$$ $$
 \times
 \exp
 [
  i\omega_1 (t_1 - t_3)
  +
  i\omega_2 (t_2 - t_3)
 ]
 \la
  \delta I(t_1) \delta I(t_2) \delta I(t_3)
 \ra
$$
is given  by an expression
$$
 S_3(0, 0)
 =
 12
 \frac{e}{RL^2}
 \int d\eps
 \int dx \int dx'\,
 [
   1 - 2f(\eps, x)
 ]
 K(x, x')
$$ \begin{equation}
 \times
 f(\eps, x')
 [
   1 - f(\eps, x')
 ].
 \label{Q-S3.vs.f}
\end{equation}
The first, second, and the third terms in the right-hand side of
Eq. (\ref{D-dI-4}) allow 6, 12, and 3 inequivalent permutations of
indices, respectively. Hence the low-frequency Fourier transform
of fourth cumulant of current
$$
 S_4(\omega_1, \omega_2, \omega_3)
 =
 \int d(t_1 - t_4)
 \int d(t_2 - t_4)
 \int d(t_3 - t_4)
$$ $$
 \times
 \exp
 [
  i\omega_1 (t_1 - t_4)
  +
  i\omega_2 (t_2 - t_4)
  +
  i\omega_3 (t_3 - t_4)
 ]
$$ $$
 \times
 \la:
  \delta I(t_1) \delta I(t_2)
  \delta I(t_3) \delta I(t_4)
 :\ra
$$
may be written in a form
\begin{equation}
 S_4(0, 0, 0)
 =
 6S_{4-1} + 12S_{4-2} + 3S_{4-3},
 \label{Q-S4-sum}
\end{equation}
where
$$
 S_{4-1}
 =
 -16
 \frac{e^2}{RL^3}
 \int d\eps
 \int\limits_0^L dx
 \int\limits_0^L dx_1
 K(x, x_1)
$$ $$
 \times
 f(\eps, x_1) [1 - f(\eps, x_1)]
$$ \begin{equation}
 \times
 \int\limits_0^L dx_2
 K(x, x_2)
 f(\eps, x_2) [1 - f(\eps, x_2)],
 \label{Q-S4-1}
\end{equation}
$$
 S_{4-2}
 =
 8
 \frac{e^2}{RL^3}
 \int d\eps
 \int\limits_0^L dx
 [
   1 - 2f(\eps, x)
 ]
$$ $$
 \times
 \int\limits_0^L dx_1
 K(x, x_1)
 [
   1 - 2f(\eps, x_1)
 ]
$$ \begin{equation}
 \times
 \int\limits_0^L dx_2
 K(x, x_2)
 f(\eps, x_2) [1 - f(\eps, x_2)],
 \label{Q-S4-2}
\end{equation}
and
$$
 S_{4-3}
 =
 8
 \frac{e^2}{RL^3}
 \int d\eps
 \int\limits_0^L dx_1
 [
   1 - 2f(\eps, x_1)
 ]
$$ $$
 \times
 \int\limits_0^L dx_2
 [
   1 - 2f(\eps, x_2)
 ]
$$ \begin{equation}
 \times
 \int\limits_0^L dx
 K(x_1, x)
 K(x_2, x)
 f(\eps, x)[ 1 - f(\eps, x) ].
 \label{Q-S4-3}
\end{equation}
\begin{figure}
\epsfxsize8cm \centerline{\epsffile{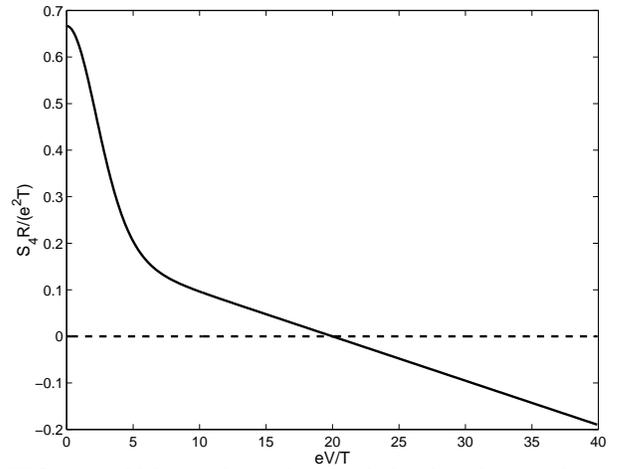}} \caption{
  Voltage dependence of the fourth cumulant of current for a long
  diffusive contact
} \label{fig_2}
\end{figure}
\noindent
The distribution function inside the contact is given by
an expression\cite{Nagaev-92}
\begin{equation}
 f(\eps, x)
 =
 \left(
  1 - \frac{x}{L}
 \right)
 f_0(\eps + eV/2)
 +
 \frac{x}{L}
 f_0(\eps - eV/2),
 \label{Q-f}
\end{equation}
where $f_0(\eps)$ is the Fermi distribution function. The
integration in Eqs. (\ref{Q-S3.vs.f}) - (\ref{Q-S4-3}) is easily
performed. One obtains that the third cumulant is equal
\begin{equation}
 S_3
 =
 \frac{1}{15}
 \frac{e}{R}
 \frac{
   eV
   \cosh(eV/T)
   +
   12T
   \sinh(eV/T)
   -
   13eV
 }{
   \cosh(eV/T) - 1
 }.
 \label{Q-S3.vs.V}
\end{equation}
This is essentially the same expression that was obtained by
Gutman and Gefen\cite{Gutman-02} using the nonlinear
$\sigma$-model. It gives
\begin{equation}
 S_3
 =
 \frac{1}{15}
 \frac{e^2 V}{R}
 \label{Q-S3-high}
\end{equation}
at high voltages in agreement with Lee et al.\cite{Lee-95} and
\begin{equation}
 S_3
 =
 \frac{1}{3}
 \frac{e^2 V}{R}
 \label{Q-S3-low}
\end{equation}
in the low-voltage or high-temperature limit.\cite{Gutman-02}

The fourth cumulant of current is given by an expression
$$
 S_4
 =
 -\frac{1}{420}
 \frac{e^2}{R}
 \frac{1}{
   \sinh^3(eV/2T)
 }
 \left[
   eV
   \cosh
   \left(
    \frac{3eV}{2T}
   \right)
 \right.
$$ $$
 \left.
   -
   20T
   \sinh
   \left(
    \frac{3eV}{2T}
   \right)
   -
   313eV
   \cosh
   \left(
    \frac{eV}{2T}
   \right)
 \right.
$$ \begin{equation}
 \left.
   +
   684T
   \sinh
   \left(
    \frac{eV}{2T}
   \right)
 \right].
 \label{Q-S4.vs.V}
\end{equation}
It gives
 \begin{equation}
 S_4
 =
 -\frac{1}{105}
 \frac{e^3 V}{R}
 \label{Q-S4-high}
\end{equation}
in the high-voltage limit in agreement with the results of Lee et
al.\cite{Lee-95}. In equilibrium at $V = 0$ only the last term in
Eq. (\ref{Q-S4-sum}) is nonzero and
\begin{equation}
 S_4
 =
 \frac{2}{3}
 \frac{e^2T}{R}
 \label{Q-S4-low}
\end{equation}
This result seems to be new. The positive sign of $S_4$ in
equilibrium is in a qualitative agreement with the shape of
distribution function of current numerically obtained by Belzig
and Nazarov.\cite{Belzig-01} The behavior of $S_4(V)$ is shown in
Fig. 2. The fourth cumulant changes its sign at $eV \approx 20T$,
hence the distribution of current changes from super-Gaussian to
sub-Gaussian as the voltage increases.

It has been proposed recently\cite{Levitov-01} that the third
cumulant of noise may be used for determining the effective charge
of quasiparticles. In other words, it may give the same
information as the shot noise even at low voltages $eV \ll T$.
Since the fourth cumulant is nonzero  even at zero voltage, one
may obtain a nontrivial information about the system  from
measurements of equilibrium fluctuations. Measurements of
fluctuations of noise power, which are termed "the second spectral
density" and are closely related with the Fourier transform of the
fourth cumulant, \cite{Kogan-book} were successfully implemented
by a number of authors who studied the $1/f$
noise.\cite{Garfunkel-90,Israeloff-91,Parman-92}

In summary, we have shown that higher cumulants of current in
diffusive-metal contacts may be described by the semiclassical
Boltzmann - Langevin theory. The semiclassical 
fluctuations appear to be non-Gaussian both in equilibrium and at 
high voltages.\cite{comment} The present approach may be also applied 
to semiclassical non-diffusive systems. 
In this case, one has to take into account also higher-order local
correlators of extraneous sources.

The author acknowledges a helpful discussion with Sh. M. Kogan.
This work was supported by Russian Foundation for Basic Research,
grant 01-02-17220, and by an INTAS Open grant,  and by an INTAS
Open grant 2001-1B-14.

\bigskip
\centerline{\bf
                        APPENDIX
}

\bigskip
The third central moment of $\delta J^{ext}$ is easily found from
the same considerations as the second moment. Much like the
second-order correlator, it is local in space. As the collision
integral involves only single-particle transitions between pairs
of states, the triple correlator $\la\delta J^{ext}(\bp_1)\delta
J^{ext}(\bp_2)\delta J^{ext}(\bp_3)\ra$ is nonzero only if at
least two of these states coincide. We also make use of the fact
that the third-order correlator of a scattering flux from one
state to another, which is assumed to be Poissonian, is
proportional to the average scattering rate between these states.
The sign of the contribution from each scattering process should
be determined by multiplying the signs of occupancy changes in
states $\bp_1$, $\bp_2$, and $\bp_3$ when the corresponding
scattering event takes place. Hence the correlator may be written
in a form
$$
 \la
   \delta J^{ext}(\bp_1, \br_1, t_1)
   \delta J^{ext}(\bp_2, \br_2, t_2)
   \delta J^{ext}(\bp_3, \br_3, t_3)
 \ra
$$ $$
 =
 \delta(\br_1 - \br_2)
 \delta(\br_2 - \br_3)
 \delta(t_1   -   t_2)
 \delta(t_2   -   t_3)
$$ $$
 \times
 \Biggl\{
   \delta_{\bp_1\bp_2}
   \delta_{\bp_2\bp_3}
   \sum\limits_{\bp'}
   \left[
     J(\bp'\to\bp_1) - J(\bp_1\to\bp')
   \right]
$$ $$
   +
   \delta_{\bp_1\bp_2}
   \left[
     J(\bp_1\to\bp_3) - J(\bp_3\to\bp_1)
   \right]
$$ $$
   +
   \delta_{\bp_2\bp_3}
   \left[
     J(\bp_2\to\bp_1) - J(\bp_1\to\bp_2)
   \right]
$$ $$
   +
   \delta_{\bp_1\bp_3}
   \left[
     J(\bp_1\to\bp_2) - J(\bp_2\to\bp_1)
   \right]
 \Biggr\},
 \eqno (A1)
$$
where the scattering fluxes $J$ are given by Eq. (\ref{G-flux}).
It is easily verified that this correlator conserves the total
number of electrons at a given point.

Consider now the fourth-order {\it irreducible} correlator of
extraneous sources. The term "irreducible" means that only the
part of the correlator that cannot be decoupled into products of
second-order correlators is considered. This correlator is zero
for a Gaussian noise but is proportional to the average rate for a
Poissonian process. Using the same ideas, the irreducible
correlator may be written in a form
\mybeginwide
$$
 \la:
   \delta J^{ext}(\bp_1, \br_1, t_1)
   \delta J^{ext}(\bp_2, \br_2, t_2)
   \delta J^{ext}(\bp_3, \br_3, t_3)
   \delta J^{ext}(\bp_4, \br_4, t_4)
 :\ra
 =
$$ $$
 \delta(\br_1 - \br_2)
 \delta(\br_2 - \br_3)
 \delta(\br_3 - \br_4)
 \delta(t_1   -   t_2)
 \delta(t_2   -   t_3)
 \delta(t_3   -   t_4)
$$ $$
 \times
 \Biggl\{
   \delta_{\bp_1\bp_2}
   \delta_{\bp_2\bp_3}
   \delta_{\bp_3\bp_4}
   \sum\limits_{\bp'}
   \left[
     J(\bp'\to\bp_1) + J(\bp_1\to\bp')
   \right]
   -
   \delta_{\bp_1\bp_2}\delta_{\bp_2\bp_3}
   \left[
     J(\bp_1\to\bp_4) + J(\bp_4\to\bp_1)
   \right]
$$ $$
   -
   \delta_{\bp_1\bp_2}\delta_{\bp_2\bp_4}
   \left[
     J(\bp_1\to\bp_3) + J(\bp_3\to\bp_1)
   \right]
   -
   (
     \delta_{\bp_1\bp_3}\delta_{\bp_3\bp_4}
     +
     \delta_{\bp_2\bp_3}\delta_{\bp_3\bp_4}
   )
   \left[
     J(\bp_1\to\bp_2) + J(\bp_2\to\bp_1)
   \right]
$$ $$
   +
   \delta_{\bp_1\bp_2}\delta_{\bp_3\bp_4}
   \left[
     J(\bp_1\to\bp_3) + J(\bp_3\to\bp_1)
   \right]
$$ $$
   +
   (
     \delta_{\bp_1\bp_3}\delta_{\bp_2\bp_4}
     +
     \delta_{\bp_1\bp_4}\delta_{\bp_2\bp_3}
   )
   \left[
     J(\bp_1\to\bp_2) + J(\bp_2\to\bp_1)
   \right]
 \Biggr\}.
   \eqno (A2)
$$
\myendwide
In the diffusive limit, it is convenient to pass from extraneous
fluxes in the Boltzmann - Langevin equation to extraneous currents
in the drift-diffusion equation by means of Eq. (\ref{D-dj^ext}).
The third-order correlator of extraneous currents is of the form
$$
 \la
  \delta j_{\alpha}^{ext}(\br_1, t_1)
  \delta j_{\beta}^{ext}(\br_2, t_2)
  \delta j_{\gamma}^{ext}(\br_3, t_3)
 \ra
$$ $$
 =
 \delta(\br_1 - \br_2)
 \delta(\br_2 - \br_3)
 \delta(t_1 - t_2)
 \delta(t_2 - t_3)
$$ $$
 \times\tau^3 e^3
 \sum\limits_{\bp\bp'}
 (
   v_{\alpha}v_{\beta}v_{\gamma}
   -
   v_{\alpha}'v_{\beta}v_{\gamma}
   -
   v_{\alpha}v_{\beta}'v_{\gamma}
   -
   v_{\alpha}v_{\beta}v_{\gamma}'
 )
$$ $$
 \times
 [
   J(\bp'\to\bp) - J(\bp\to\bp')
 ].
 \eqno (A3)
$$
Though the scattering fluxes $J$ are proportional to $1/\tau$, the
bracketed difference vanishes in equilibrium and is proportional
to the anisotropic part of the distribution function in the case
of applied bias. As this part is proportional to $\tau$ in the
diffusive approximation,  Eq. (A3) is at least of the order of
$\tau^3$.

The fourth-order irreducible correlator of extraneous currents is
of the form
$$
 \la:
  \delta j_{\alpha}^{ext}(\br_1, t_1)
  \delta j_{\beta}^{ext}(\br_2, t_2)
  \delta j_{\gamma}^{ext}(\br_3, t_3)
  \delta j_{\delta}^{ext}(\br_4, t_4)
 :\ra
$$ $$
 =
 \frac{16}{15}
 e^2 l_{imp}^2 \sigma
 \delta(\br_1 - \br_2)
 \delta(\br_2 - \br_3)
 \delta(\br_3 - \br_4)
$$ $$
 \times
 \delta(t_1 - t_2)
 \delta(t_2 - t_3)
 \delta(t_3 - t_4)
 (
   \delta_{\alpha\beta} \delta_{\gamma\delta}
   +
   \delta_{\alpha\gamma}\delta_{\beta\delta}
   +
   \delta_{\alpha\delta}\delta_{\beta\gamma}
 )
$$ $$
 \int d\eps f(\eps,\br_1)[1 - f(\eps,\br_1)],
 \eqno (A4)
$$
where $\sigma$ is the conductivity. Hence the fourth-order
correlator is also of the order $\tau^3$.

\end{multicols}
\end{document}